# Localization of Epileptic Seizure Focus by Computerized Analysis of fMRI Recordings

-Original Research-

Rasoul Hekmati[1*], Robert Azencott [1], Wei Zhang[2,3], Zili D. Chu[2,4], Michael J. Paldino[2]

1. Department of Mathematics, University of Houston, Houston, Texas, United States of America
2. Department of Radiology, Texas Children's Hospital, Houston, Texas, United States of America
3. Outcomes & Impact Service, Texas Children's Hospital, Houston, Texas, United States of America
4. Department of Radiology, Baylor College of Medicine, Houston, Texas, United States of America

\* Email: rhekmati@math.uh.edu

## Abstract

By computerized analysis of cortical activity recorded via fMRI for pediatric epilepsy patients, we implement algorithmic localization of epileptic seizure focus within one of eight cortical lobes. Our innovative machine learning techniques involve intensive analysis of large matrices of mutual information coefficients between pairs of anatomically identified cortical regions. Drastic selection of pairs of regions with significant inter-connectivity provide efficient inputs for our Multi-Layer Perceptron (MLP) classifier. By imposing rigorous parameter parsimony to avoid over fitting we construct a small size MLP with very good percentages of successful classification.

### Author summary

The localization of epileptic focus involves discrimination between 8 classes of epileptic patients. We construct and train an efficient artificial neural network, of MLP type (Multi-Layer Perceptron) to solve this classification task. One originality of our approach is to rigorously construct a parsimonious MLP architecture of minimal size in order to achieve robust classification based on a moderate number of patients. To this end we develop an algorithm to radically reduce the number of features extracted from fMRI recordings. This feature selection is based on comparisons of discriminating power between cortex regions. To increase robustness, we take into account the errors of estimation for mutual information coefficients, and we use "German Tank Estimates" to mitigate the moderate size of our data set. The performance is high and the intra/inter region connections obtained from resting state fMRI show a strong association to clinical diagnosis of seizure onset zone.

## Introduction

Epilepsy is a common brain disorder that affects 6 out of 1000 children in the United States [1, 2]. Although medication is the mainstay of treatment for these patients, seizures are resistant to anti-epileptogenic drugs in approximately 30% of children. Such medically-refractory epilepsies are frequently "focal", defined as originating habitually from a relatively localized region of the brain. In appropriately selected patients, resection of this localized seizure origin can be curative. However, surgical outcomes remain highly variable, even in optimal candidates [3-6]. Emerging studies motivated by these inconsistencies have demonstrated that pediatric focal epilepsy is not a localized disorder, but rather a disease of large-scale, distributed neural networks [7]. Furthermore, rather than resulting from abnormal activity generated by one or more abnormal foci, the origin of seizures occurs within a functionally and anatomically connected set of brain regions – a seizure network [8]. Failure to sufficiently resect or disconnect the seizure network may be associated with continued seizures, thereby contributing to surgical failure[9]. At the current time, there is no modality that can accurately map the extent of the seizure network.

Advances in MRI combined with mathematical network approaches have demonstrated great potential to study brain networks non-invasively in a variety of populations, including children with focal epilepsy [10, 11]. Resting-state functional MRI is one method by which connectivity in the brain can be measured. This sequence quantifies the blood oxygen level dependent (BOLD) signal over time as an indirect measure of neuronal activity[12]. Functionally connected elements of the brain exhibit similar spontaneous BOLD signal fluctuations at rest[12], which enables one to capture interactions between brain regions *in vivo*. Within the network framework, the brain is represented as a graph, a mathematical construct consisting of a set of nodes connected by edges[13]. The nodes are defined as anatomical regions of the brain; the edge between each pair of nodes is estimated as the magnitude of association between their BOLD time series. Studies based on such resting state functional network constructs have consistently shown global aberrations of the epileptic brain from normal; they have also demonstrated that network features have the potential to predict clinically-relevant aspects of brain function in children with epilepsy[14-18]. However, little data exists regarding the use of neuroimaging-defined network models to identify seizure networks.

The goal of this study was to assess the potential of imaging-defined networks to depict seizure networks. Specifically, we trained an artificial neural network (ANN) to identify the location of seizure origin given access only to features derived from a resting state functional network. We measured the accuracy of the learning algorithm against the lobe of seizure origin determined at multi-disciplinary epilepsy surgery conference. We selected this reference standard - which relies on the consensus of multiple objective tests as well as the clinical expertise from multiple disciplines - as an accurate and reliable assessment of the lobe of seizure origin[19, 20]. Identification of the lobe of seizure origin can be considered an initial step toward the ultimate validation of the potential for resting state functional networks to visualize seizure networks in children with focal epilepsy.

## 1.1 Materials and methods

*Patients:*

The HIPAA-compliant study was approved by a local institutional review board. Informed consent was waived for this retrospective study. Patient medical records were retrospectively reviewed to identify patients with the following inclusion criteria: 1) pediatric age group (21 years of age or younger); 2) a clinical diagnosis of focal epilepsy; 3) an available 3 Tesla MR imaging of brain, including a rs-fMRI sequence; and 4) lobe of seizure origin identified at multidisciplinary epilepsy conference. This determination relies on the consensus of multiple objective tests (including MRI and EEG) as well as the clinical expertise from multiple disciplines (including Neurology, Neurosurgery, and Neuroradiology) [19, 20]. The above defined cohort was refined by applying the following exclusions: 1) any brain operations performed prior to the MR imaging. Images were performed from January 2012 to December 2017. Forty-six patients were included. Lobe diagnoses for the cohort are presented in Table 1.

| Lobe | Frequency |
|---|---|
| Group 1 - Right Frontal (RF) | 10 |
| Group 2 - Left Frontal (LF) | 8 |
| Group 3 - Right Temporal (RT) | 7 |
| Group 4 - Left Temporal (LT) | 11 |
| Group 5- Others: Right & Left Parietal + Right & Left Occipital (C5) | 6 |

**Table 1:** Lobe diagnosis of the cohort

The 5 classes are not perfectly disjoint because 4 patients were ambiguously diagnosed by clinicians and thus belong to 2 classes simultaneously (RTC5,RFRT,LFLT,LTRF).

*MR Imaging:*

All imaging was performed on a 3 Tesla Achieva system (Philips, Andover, Massachusetts) with a 32-channel phased array coil. The following sequences were obtained: 1. Structural images: sagittal volumetric T1-weighted images (repetition time (TR)/echo time (TE): 7.2 ms/2.9 ms; 1 acquisition; Flip angle: 7 degrees, Inversion time: 1100 ms; field of view (FOV): 22 cm; voxel size (mm): 1 x 1 x 1)). 2. Resting state fMRI: axial single-shot echo planar imaging (EPI) fMRI (TR/TE (ms): 2000/30; flip angle: 80 degrees; 1 acquisition; FOV: 24 cm; voxel (mm): 3 x 3 x 3.75; 300 volumes (duration: 10 minutes)) performed in the resting state. Patients were instructed to lie quietly in the scanner with their eyes closed. All images were visually inspected for artifacts, including susceptibility and subject motion.

*Image processing and analysis:*

The processing pipeline was implemented using MATLAB scripts (version 7.13, MathWorks, Inc) in which adapter functions were embedded to execute FreeSurfer reconstruction (version 5.3.0; http://surfer.nmr.mgh.harvard.edu) and several FMRIB Software Library (FSL) suite tools [21]. Details regarding this pathway have been previously described. [17, 18] A brief summary is provided here:

*Network Node Definition:*

The reference space was created from images of one patient in our database, who had no visible abnormality and with optimal registration to MNI space [22]. Structural imaging data for each patient were aligned to a standard reference template (MNI152) using FSL's non-linear registration algorithm [21, 23]. Nodes in the network were defined on the template according to parcellation of whole-brain gray matter. First, FreeSurfer reconstruction of cerebral cortical surfaces was performed on the T1 structural image. This processing stream includes motion correction, skull stripping, intensity normalization, segmentation of white matter and gray matter structures, parcellation of the gray matter and white matter boundary, and surface deformation following intensity gradients which optimally place the gray matter/white matter and gray matter/cerebrospinal fluid borders [24, 25]. The pial and gray white surfaces were visually inspected using the Freeview software for accurate placement.

Next, a self-developed MATLAB program was applied to the FreeSurfer output to further subdivide the 148 standard gray matter region according to their surface area. During this process, each region was iteratively divided into two new regions of equal size until the surface area of each parcel (as defined on the FreeSurfer gray-white surface mesh) was less than a size threshold of 350-mm$^2$. The final parcellation contained 780 nodes (Fig 1). Each surface parcel was then converted into a volume mask of gray matter at that region to form a node on the network. All nodes defined in reference space were transformed into each individual patient's space by applying the nonlinear transformation matrix (12 degrees-of-freedom) obtained during registration.

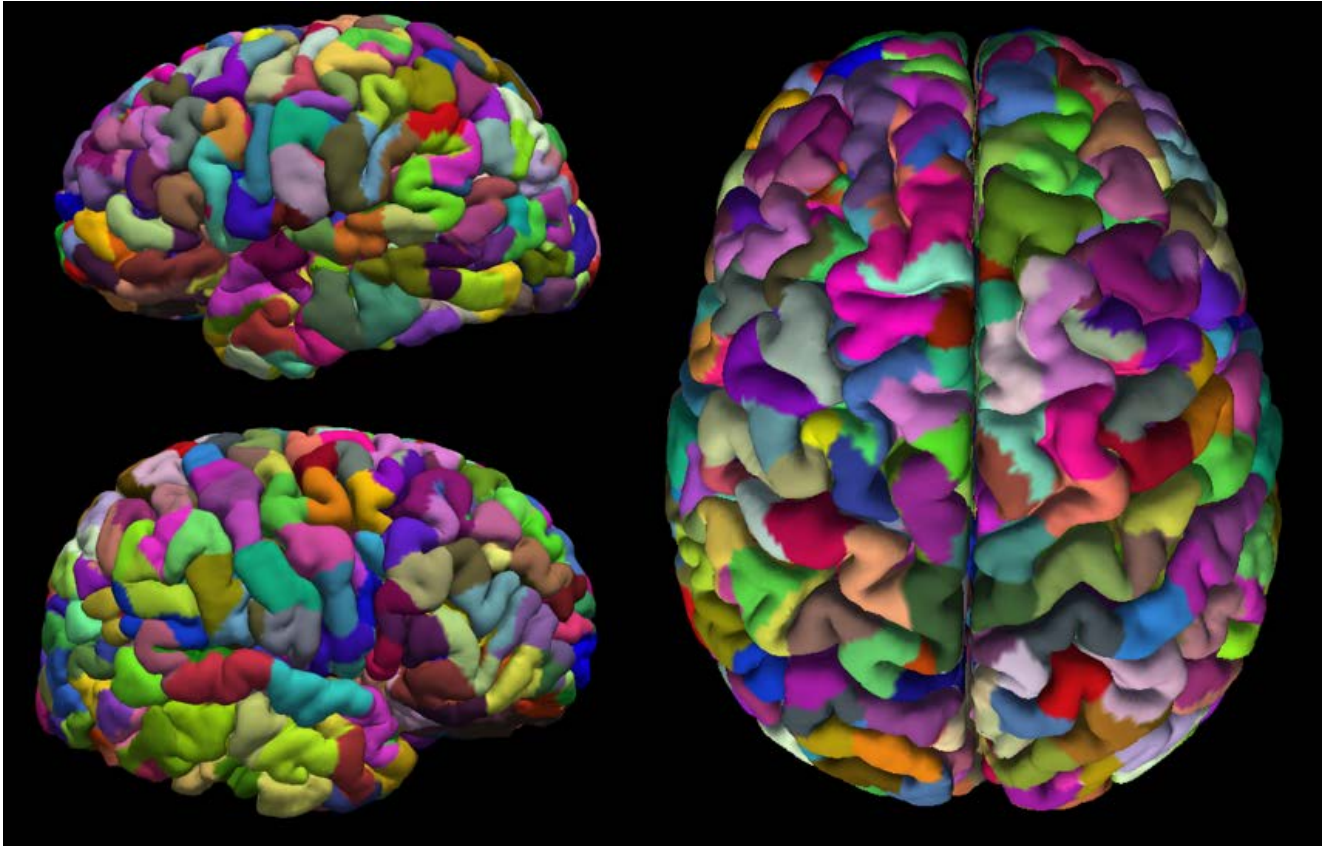

**Figure 1: Final parcellation with a size threshold of 350-mm$^2$, resulting in 780 nodes.**

*FMRI data preprocessing:*

The first 5 volumes in each resting state functional data were removed to allow magnetization to reach equilibrium. Standard preprocessing and independent component analysis (ICA) of the functional data sets was performed using FSL MELODIC [21], consisting of motion correction, interleaved slice timing correction, brain extraction, spatial smoothing with a Gaussian kernel full width at half maximum of 5 mm, and high pass temporal filtering equivalent to 100 seconds (0.01 Hz). Noise related to motion and other physiologic nuisance was addressed according to an independent component analysis technique [26]. Nonsignal components were removed manually by an expert operator with 6 years of experience using independent component analysis in this patient population. Although the optimal strategies for noise removal in fMRI are debatable [27, 28], an independent component analysis was selected because it has been shown to minimize the impact of motion on network metrics while, at the same time, decreasing the loss of temporal degree of freedom and preserving the signal of interest across a variety of resting-state datasets [28], Affine boundary-based registration as implemented in FSL FLIRT was then used to align the pre-processed functional image volumes for each patient to that individual's structural T1 dataset using linear registration. The inverse transformation matrix was calculated in this step and subsequently used to transform all masks from structural to functional space. Mean BOLD-signal time series were then computed for each node.

## 2. Mutual information between time series

### 2.1 Mutual information:

Mutual Information $MI(X,Y)$ between two random variables $X$ and $Y$ quantifies the amount of information brought by the knowledge of $X$ to the prediction of $Y$. This quantification is actually symmetric in $X$ and $Y$, and is based on the concept of conditional entropy. When $X$ and $Y$ can take only a finite number of values denoted $x_k$ and $y_m$, denote $p_k$, the entropy $H(X)$ of $X$ is given by

$$p_k = Pty(X = x_k); \quad q_m = Pty(Y = y_m); \quad r_{k,m} = Pty(X = x_k, Y = y_m)$$

Then the entropies of $X, Y$, and $Z = (X, Y)$ are given by

$$H(X) = -\sum_k p_k \log(p_k) \geq 0; \quad H(Y) = -\sum_m q_m \log(q_m);$$

$$H(X,Y) = -\sum_{k,m} r_{k,m} \log(r_{k,m})$$

The Mutual Information between $X$ and $Y$ is given by

$$MI(X,Y) = MI(Y,X) = H(X) + H(Y) - H(X,Y) \geq 0$$

and $MI(X,Y) = 0$ when $X$ and $Y$ are independent.

When $X$ and $Y$ have a bivariate Gaussian distribution, the mutual information $MI(X,Y)$ is computable as an explicit increasing function of the squared correlation,

$$MI(X,Y) = -\frac{1}{2}\log(1 - cor(X,Y)^2$$

In the non-Gaussian case, this formula is not valid, but mutual information has major advantages over correlations. Indeed one always has $MI(X,Y) = MI(f(X), f(Y))$ for any invertible but not necessarily linear function $f$. This is a key point here since at time "$t$" our fMRI data provide image intensities $Y_i(t)$ for each parcel $i$ which are linked to the unknown blood level $BL_i(t)$ by some (unknown) and non linear function: $Y_i(t) = f(BL_i(t))$. To study brain fMRI time series, Mutual Information between time series is a much more stable quantity than the often used absolute values of correlations.

So we expect strong values of $MI_{i,j} = MI(Y_i, Y_j)$ to detect the frequent presence of simultaneous high blood levels for cortex parcels $CP_i$ and $CP_j$, a phenomenon indicative of strong neuronal interactivity between $CP_i$ and $CP_j$. Since each observed time series $Y_i(t)$ involves 295 observations of a continuous random variable $Y_i$, we discretize each $Y_i$ by first computing its quantiles at [20%, 40%, 60%, 80%]. This defines 5 intervals $[Z_1, ..., Z_5]$ with midpoints $[z_1, ..., z_5]$, and whenever $Y_i(t)$ falls in $Z_k$, we replace $Y_i(t)$ by $Y_i(t) = z_k$. Then we approximate $MI_{i,j}$ by $MI(\widehat{Y}_i, \widehat{Y}_j)$. To evaluate the accuracy of our mutual information estimates, we have applied a variant of the "Jackknife resampling" technique. For $MI_{i,j}$ values lying between the 20% and 80% quantiles of the more than $1.5 10^6$ mutual information coefficients computed on our data set, our error of estimation on $MI_{i,j}$ had typical size less than $10\% \times MI_{i,j}$.

The use of 20 bins instead of 5 bins to generate the discretized $Y_i(t)$ tends to double the average errors of estimation for the $MI_{i,j}$, as can be expected from theory. Indeed our exploration of more detailed discretization, and even optimized discretization, has shown that using 5 bins of equal frequency 20% for each $Y_i$ was quite efficient in our context.

## 2.2 Mutual information between pairs of cortex sub-regions:

For each patient, there are roughly 780 parcels of size 350 mm$^2$ and for each parcel, one time series with 295 recordings. We characterize the cortex connectivity of each patient by the symmetric $780 \times 780$ mutual information matrix $MI_{i,j}$. The coefficients of this MI matrix provide a very large number of features $\cong 780^2/2$ [29]. To reduce the number of features we have developed and implemented an innovative multi-scale analysis of these MI matrices.

For any cortex subregion $A$, denote $L(A)$ the list $i(1), \ldots, i(n_A)$ of $n_A$ cortex parcels contained in $A$. For any two cortex subregions $A$ and $B$, generate the set $S(A, B)$ of all $n_A$ by $n_B$ mutual information coefficients $MI_{i,j}$ with $i$ in $L(A)$ and $j$ in $L(B)$. We then define the mutual information $MI(A, B) = MI(B, A)$ as the 75% quantile of the list $L(A, B)$. When $MI(A, B)$ is high, at least 25% of cortex parcels pairs $i$ in $L(A)$ and $j$ in $L(B)$ have high $MI_{i,j}$ are expected to have strong neural interaction. In our definition of $MI(A, B)$ we could have used other quantiles of $S(A \times B)$. After methodical exploration of other choices ranging from 60% to 90% quantiles, we have concluded that 75% quantiles were most efficient for our classification task.

We then compute the coefficients $MIreg(m,n) = MI(REG_m, REG_n)$ for any two regions in our list of 148 anatomically defined cortex regions to obtain for each patient a symmetric matrix $MIreg$ of "Cortex Regions Interactions". $MIreg$ has size $148 \times 148$, and each one of its (148) (149)/2 = 11026 distinct coefficients will provide potential input features for our patients classification task.

We have also studied the symmetric $10 \times 10$ symmetric matrices $MIlob$ of mutual information between cortex lobes. Fig 2 displays (for only one patient) the matrices $MIreg$ and $MIlob$.

## 3. Automatic classification by multilayer perceptron

### 3.1 Parameters parsimony constraints for MLPs:

The restricted number of diagnosed patients in our data set imposes a strong parsimony for the number of parameters to be "learned" by any *robust* classifier [30, 31]. So we have studied intensively the performances of patient classification into 5 classes by Multilayer Perceptron

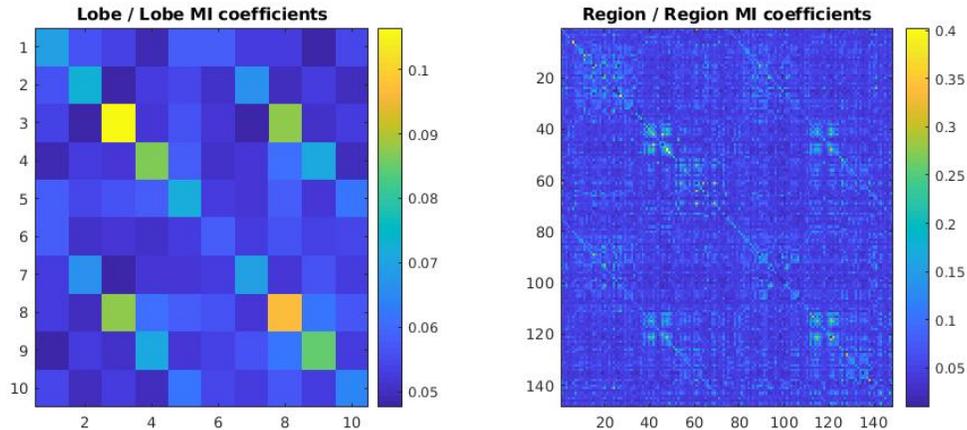

**Figure 2:** Mutual Information Matrices for 148 cortex Regions (right-hand fig.) and for 10 cortex Lobes (left-hand fig.). Matrix coefficients are displayed via 5 color heat map

(MLPs) of small size. For our MLPs, we have imposed a strong reduction of the number of unknown "weights" to be calibrated by automatic learning. Indeed deep theoretical results of Vapnik ([32]) indicate that this is a good recipe to perform stable automatic learning on a moderately large training set.

Feed forward neural networks (such as MLPs) are known to exhibit better generalization capacity and less overfitting when their architecture involves fewer weights. But one still wants the MLPs to achieve good classification performance, which often requires to increase the number of weights. Our approach to implement an ideal balance between these two criteria has involved a meticulous algorithmic selection of a small number of highly discriminating input features for our MLPs. Let "$f$" be the number of selected input features and "$k$" be the number of patients classes. Our MLP classifiers involve 4 successive layers of artificial neurons: an input layer of size $f$, a hidden layer of size $h$, an "internal" output layer of size $k$, and finally a "softmax" layer of size $k$ which transforms the $k$ internal outputs into $k$ decision probabilities with sum = 1. The number $W$ of free weights and thresholds (also called free parameters) of such an MLP is given by: $W = h(f + k + 1) + k$.

Call $S$ the size of the training set. Each training case provides $k$ precise values for the $k$ decision probabilities computed by the softmax layer, but by construction these $k$ probabilities have sum = 1. Each case hence yields $(k - 1)$ equations which must be verified by the $W$ free parameters. The training set thus provides $(k - 1) \times S$ nonlinear equations to be satisfied by $W$ unknowns. To avoid overfitting one should "ideally" have

$W < (k-1) \times S$, which yields the following dimensional constraint on "$h$" and "$f$",

$$\text{Parameters Parsimony Constraint} \quad h(f+k+1) < (k-1)S - k$$

Here we have $k = 5$ patient classes and $S = 45$ because we use leave-one-out training. So the parameters parsimony constraint on $(h, f)$ becomes

$$\text{Parameters Parsimony Constraint} \quad h(f+6) \leq 175$$

### 3.2 Automatic learning, implementation:

Once the $f$ input features are selected as well as the hidden layer size $h$, with the preceding parsimony constraint, we finalize the MLP architecture by selecting the "Sigmoid Response" function for all individual neurons in the 2nd and 3rd layer. For ambiguously diagnosed patients belonging to two classes $CL(p)$ and $CL(q)$, we train the network to output probabilities 1/2 for each one of these two decisions. Automatic learning was driven by standard gradient descent, with "Cross Entropy" as Loss Function. Indeed for classification tasks, minimizing Cross Entropy is generally more efficient than minimizing Mean Squared Error.

Since our training set is of moderate size, we implement the classical Leave-One-Out technique. Namely, one eliminates one patient from the training set before performing automatic learning; then one evaluates a 0 or 1 performance for automatic classification of the eliminated patient. Final performance is the average of these 0 or 1 performances over all possible choices of the left-out patient.

# Input selection for minimal size MLP classifiers

## 4.1 Discriminating power of generic features:

4. We want to select "features" discriminating between 5 classes of patients $CL(1),\ldots CL(5)$ of sizes s(1) ... s(5). This requires to solve at least 10 basic tasks of the type "discriminate $CL(p)$ versus $CL(q)$". Consider any feature F computable from fMRI recordings, and hence providing for each patient $PAT$ a number $F(PAT)$. For each class $CL(p)$, the $s(p)$ patients $[PAT\ 1, PAT\ 2, \ldots]$ of $CL(p)$ generate a list $V(p)$ of $s(p)$ values $F1 = F(PAT1), F2 = F(PAT2)$, ... . The probability distribution $\mu_p(F)$ of $[F1, F2, \ldots]$ is unknown, and difficult to estimate if $s(p)$ is not very large. For feature efficiency evaluation only, we roughly approximate $\mu_p(F)$ by a uniform distribution on an unknown interval $J_p(F)$. A classical algorithm to estimate $J_p(F)$ is the "German Tank Estimate" (see appendix), which computes first the min $m(p)$ and the max $M(p)$ of the list $V(p)$, and then extends adequately the interval $[m(p), M(p)]$.

When $J_p(F)$ and $J_q(F)$ are nearly disjoint intervals, we naturally consider that the classes $CL(p)$ and $CL(q)$ are strongly separated by feature $F$. To quantify this notion, define the *separability* $sep(J,J')$ of two intervals J and J' with intersection $I = J \cap J'$ by

$$sep(J,J') = \min(\frac{|J - I|}{|J|}, \frac{|J' - I|}{|J'|})$$

Where $K$ is the length of interval $K$. Then $0 \leq sep(J,J') \leq 1$, and values of $sep(J,J')$ close to 1 mean that $J, J'$ are nearly disjoint. Now define $DIS_{p,q}(F) = sep(J_p(F), J_q(F))$ as the *discriminating power* of feature $F$ to differentiate between classes $CL(p)$ and $CL(q)$.

### 4.2 Highly discriminating region/region interactions

To select a small but efficient set of "$f$" input features for our MLP classifier, we start from the (148)(149)/2 mutual information coefficients $MIreg(m,n)$ at the cortex regions "scale". For each pair $m \leq n$, the "regions based" feature $F_{m,n} = MIreg(m,n)$ provides for each patient an indicator of the neuronal interaction level between cortex regions $REG_m$ and $REG_n$.

Fix any two distinct classes $CL(p), CL(q)$. For all $(m,n)$ with $1 \leq m \leq n \leq 148$ we compute the power $POW(m,n) = DIS_{p,q}(F_{m,n})$ of feature $F_{m,n}$ to discriminate $CL(p)$ vs $CL(q)$. Then we rank the features $F_{m,n}$ by decreasing $POW(m,n)$, and select only the top two $F_{m,n}$ having highest discriminating power $POW(m,n)$.

### 4.3 Two sets of highly discriminating region/region interaction levels:

#### 4.3.1  20 strongly discriminating inter-connectivity coefficients:

For each one of the 10 pairs $CL(p), CL(q)$ with $1 \leq p < q \leq 5$, the preceding selection algorithm selects two regions $Reg_m, Reg_n$. This generates a set of 20 pairs of cortex regions $REG_m, REG_n$ with strongly discriminating coefficients $MIreg(m,n)$. These 20 pairs of regions are displayed on a 3D cortex map in Fig 3. To respect the parameter parsimony constraint, we keep only a set $InterCC$ of 18 of these Inter-Connectivity Coefficients, to be used below as a set of 18 inputs features $F_{m1,n1}, \ldots, F_{m18,n18}$ for our first MLP classifier.

**Regions with high inter-discrimination Power**

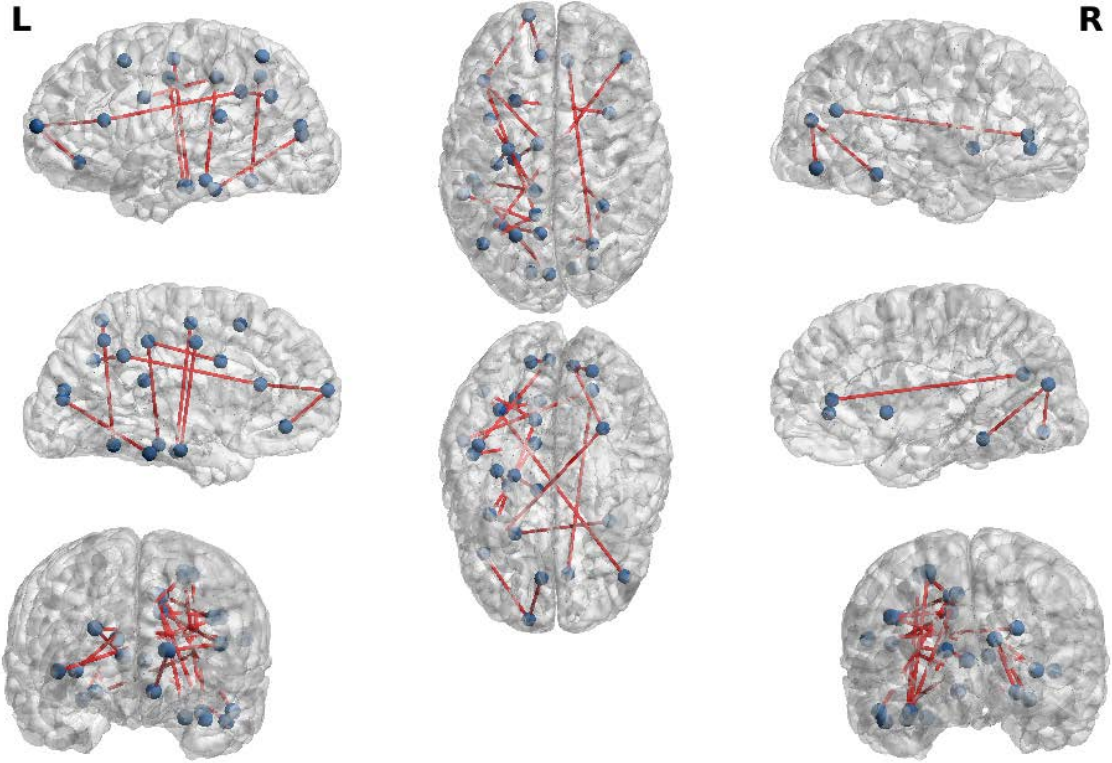

Figure 3: 20 pairs of cortex regions with strongly discriminating inter-connectivity coefficients

### 4.3.2 21 strongly discriminating intra-connectivity coefficients:

The intra-connectivity coefficient $MIreg(m, m)$ of cortex region $Reg_m$ is an indicator of the level of neuronal interactions within $Reg_m$. The selection algorithm outlined above can detect the regions $Reg_m$ for which the feature $F_{m,m} = MIreg(m, m)$ has strong power $POW(m, m)$ to discriminate between at least two classes $CL(p), CL(q)$. We have thus selected a set of 21 such cortex regions, displayed on a 3D cortex map in Fig 4. This provides 21 Intra-Connectivity Coefficients, from which we keep only a set $IntraCC$ of 18 coefficients to be used below as input features of our second MLP classifier.

**Top 3% significant regions with high intra-discrimination.**

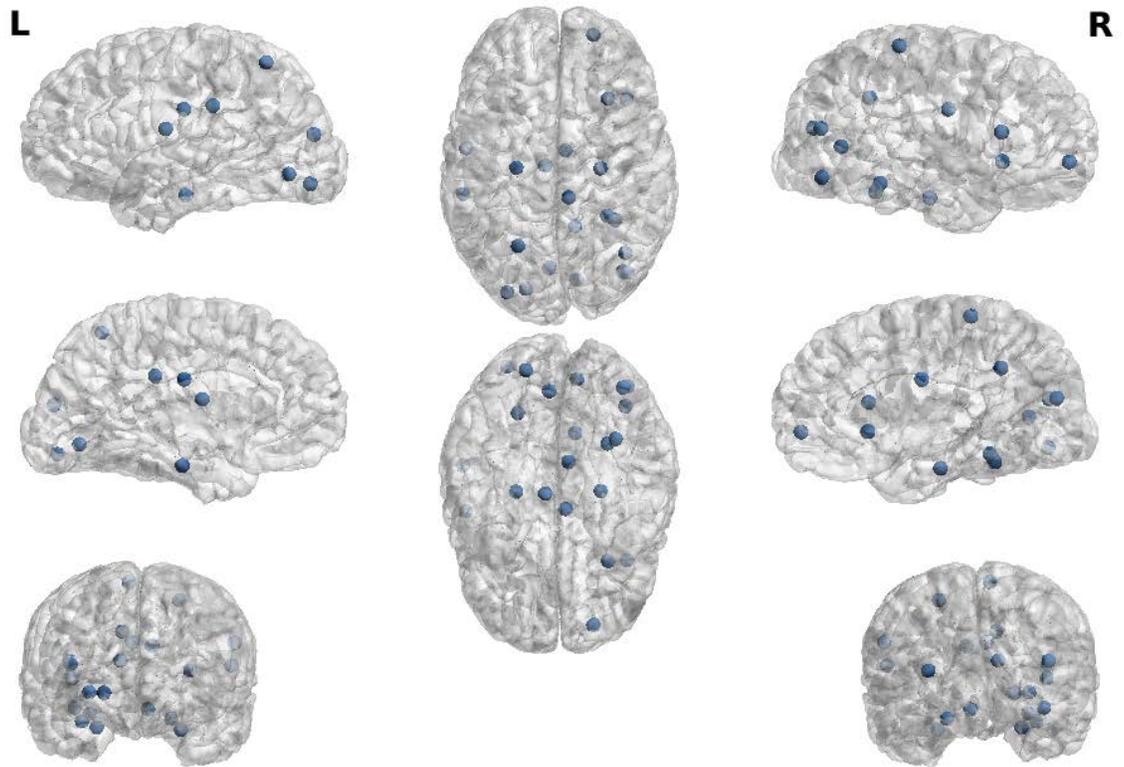

Figure 4: 21 Cortex Regions with strongly discriminating Intra-Connectivity coefficients

## *Examples of discriminating powers reached by inter- and intra-connectivity coefficients:*

**Example: ClassRT vs ClassLF.** To discriminate between ClassRT and ClassLF the most efficient Intra-Connectivity Coefficients reach the three top power values [0.8, 0.58, 0.5], and correspond to the 3 cortex regions:

"rh-S-subparietal","lh-G-oc-temp-lat-fusifor","rh-G-oc-temp-lat-fusifor"[33].

**Highest discriminating powers:** For each one of the 10 basic discrimination tasks $CL(p)$ vs $CL(q)$ the highest discriminating power $\pi(p,q)$ reached for this task among all the connectivity coefficients $MIreg(m,n)$, is a crude indicator of "how well" the task may be solved by an efficient classifier.

In Fig 5 we display in red the ten $\pi(p,q)$ values reached by Inter-Connectivity Coefficients and in blue the $\pi(p,q)$ values reached by Intra-Connectivity Coefficients. Clearly the red values dominate the blue values, and the most "difficult" task is ClassRF vs Classless.

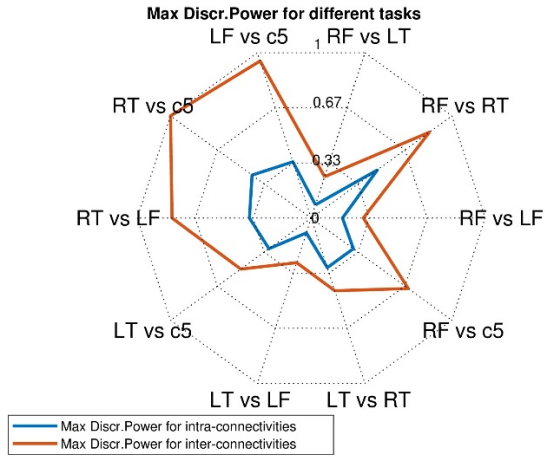

Figure 5: Highest powers reached for 10 basic discrimination tasks

5.

# Patients classification by minimal MLP with highly discriminating inputs

## 5.1 MLP classifier based on 18 inter-connectivity coefficients

The set $InterCC$ of 18 inter-connectivity coefficient $MIreg(m,n)$ selected above provides the $f = 18$ inputs of our first MLP classifier with an architecture as out- lined in section 4, for which we impose a hidden layer size $h = 7$. The 4 layers of this MLP have then sizes (18, 7, 5, 5). Our parameter parsimony equation becomes $h(f + 6) = 168 < 175$, and is hence correctly verified. This ensures a reasonable robustness of the MLP training results.

After training this small size MLP as described in subsection 4.2, we evaluate its performances by leave-one out technique. The global percentage of successful patients classifications over all 5 patients classes was 89% $\pm$ 2.1%. In the following table 2, we display the percentage of successful classifications within each patients class. Most confusions occur for discrimination between ClassRF and ClassLF.

| Patients Class | RF | LT | RT | LF | [RP,LP,RO,LO] |
|---|---|---|---|---|---|
| % Success per class | 82% | 90.5% | 100% | 87% | 90% |

Table 2

Automatic classification based on 18 strongly discriminating inter-connectivity coefficients at cortex regions scale

**5.2 MLP classifier based on 18 intra-connectivity coefficients**

The set $IntraCC$ of 18 intra-connectivity coefficients $MIreg(m,m)$ selected above provides $f = 18$ inputs for our second MLP classifier which has again a 4 layers architecture with dimensions (18, 7, 5, 5) (see section 3) and hence satisfies our parsimony equation as seen above. After training this second MLP (see subsection 3.2), performances by evaluated by leave-one out technique show a global percentage of successes of 88% $\pm$ 2.1%. Table 3 displays percentages of successes within each patients class. Results are slightly weaker than for our first MLP, particularly for ClassRF and ClassLF. Discrimination between these two classes generate most errors of classification.

| Patients Class | RF | LT | RT | LF | [RP,LP,RO,LO] |
|---|---|---|---|---|---|
| % Success per class | 70% | 80% | 100% | 87% | 100% |

Table 3

Automatic classification based on 18 highly discriminating Intra-Connectivity Coefficients

## Discussion

Our goal was to explore automatic classification of fMRI data in young epileptic patients to identify the origin of epileptic seizures. Given the moderate size of our data set, one challenge was to avoid using machine learning techniques involving large numbers of parameters. We have solved this challenge by introducing a rigorous "parameters parsimony principle".

A second challenge was to identify strongly discriminating input features computable from our fMRI data. To this end we have systematically used large matrices of Mutual Information coefficients between all pairs of roughly 780 time series extracted from each patient's fMRI data. We have introduced a computable version of Mutual Information between any two cortex regions, as an indicator of neuronal interactions between these two regions. This led us to define and compute the discriminating power of all these $148^2/2$ inter-connectivity coefficients.

We have then extracted one set of 18 strongly discriminating inter-connectivity coefficients, and used them as input features for a small size Multi-layer Perceptron (MLP) with 4 layers of dimensions (18, 7, 5, 5), including a "softmax" terminal layer. After automatic training by a leave-one out technique, this MLP provided 89% successful patients classification. For comparison, a standard Support Vector Machine applied to our patients classification has yielded less than 70% successes. Our innovative development of Mutual Information between pairs of cortex regions, and of algorithmic selection of highly discriminating pairs of cortex regions has shown good capacity to extract useful and interpretable brain activity features from fMRI recordings. The small size MLP classifier we have thus constructed rigorously avoids overfitting and reaches good performance on our group of epileptic patients. To confirm these exploratory findings, we plan to test our approach on larger groups of patients.

We also designed and implemented an analysis to interpret why interactivity between two specific regions (say $A, B$) can separate two specific lobes (say $L_1, L_2$). We found out that for almost all the patients with seizure focus in $L_1$, we can find *within* $L_1$ one or more cortex regions having strong interaction with $A$ and with $B$, while *all* subregions of $L_2$ have much weaker simultaneous interactions with $A$ and $B$. This means that $A, B$ are simultaneously excited whenever specific subregions of $L_1$ become strongly active, while hyperactivity in any subregion of $L_2$ has a weak impact on the simultaneous activity of $A$ and $B$. For example, the interconnectivity between the regions $A$ =lh_G_and_S_transv_frontopol and $B$ = lh_Lat_Fls-ant-Vertical has strong discriminating power between lobes $L_1 = LF$ and $L_2 = RF$ and we can actually find two subregions $U$ and $V$ of $LF$ with high interactivities with $A$:

$$MI(U, A) = 0.22 \text{ and } MI(V, B) = 0.25$$

while for all subregions $K$ of $RF$ one has:

$$MI(K, A) < 0.12 \text{ and } MI(K, B) < 0.11$$

Although little data exists regarding the use of whole-brain functional networks, our findings are consistent with previous applications of network connectivity to invasive EEG. Wilke et al. observed a correlation of betweenness centrality, a network graph metric, with the location of network nodes whose resection was likely to result in seizure freedom[34]. Sinha et al. also validated a model of epileptogenicity based on connectivity against surgical outcomes in a cohort consisting largely of adults [35]. They observed a correlation between their index of epileptogenicity and the putative seizure onset zone; they also found areas of high epileptogenicity outside of the resection zone in several cases of unsuccessful surgery. Tomlinson et al. applied a network approach to a pediatric cohort with focal epilepsy[36]. They observed a significant increase in connectivity outside of the seizure onset zone; further, global synchrony (a measure of the overall strength of connectivity) within the field of invasive EEG could be used to classify patients with regard to seizure-free outcome after surgery. These studies not only reinforce the relevance of network features to seizure onset, they highlight the importance of broader areas of dysconnectivity (beyond the traditional zone of seizure onset) in achieving seizure freedom. Invasive monitoring in the form of electrocorticography (ECOG) or stereo EEG (SEEG) is gold standard for localization of the seizure onset zone. These modalities, however, are limited in their spatial sampling; large areas of the brain are left unexplored, leading to the potential for erroneous conclusions[37]. Optimal use of invasive monitoring therefore requires an accurate pre-test hypothesis regarding the location and extent of the epileptogenic network. It is likely, therefore, that optimal patient outcomes would result from the addition of global network approaches to standard invasive monitoring.

## Conclusion

We developed machine learning algorithms to evaluate the connectivity obtained from resting-state fMRI in terms of differentiating the lobe of seizure origin in a pediatric cohort with focal epilepsy. These findings support the potential for neuroimaging-based network constructs to depict pathophysiologically relevant features of seizure genesis. If these approaches can be tailored to identify individual elements within the seizure network, biomarkers based on functional networks may ultimately contribute to personalized management strategies in children undergoing epilepsy surgery.

**APPENDIX**: "German Tanks" estimate for uniform distributions

Let $\mu$ be a uniform probability distribution on an unknown interval $= [\,a,b\,]$. Let $X_1, \ldots, X_N$ be $N$ independent random observations having the same probability distribution $\mu$. From these $N$ observations, one can compute the classical "German Tank" (GT) estimators $(A_N, B_N)$ of $(a, b)$, which are unbiased and have minimal variance.

To compute the GT estimators, denote

$$U_N = \min(X_1, \ldots, X_N), \quad V_N = \max(X_1, \ldots, X_N),$$

One then has

$$A_N = U_N - (V_N - U_N)/(N - 1); \quad B_N = V_N + (V_N - U_N)/(N - 1);$$